\documentclass{fizik}
\usepackage{hyperref}
\hypersetup{
colorlinks=true,
urlcolor=blue,
citecolor=blue}
\usepackage[all]{xy,xypic}
\usepackage{amsfonts,amssymb,amsmath,amsgen,amsopn,amsbsy,theorem,graphicx}
\usepackage{eufrak,amscd,bezier,latexsym,mathrsfs,enumerate,multirow,xcolor}
\usepackage[utf8]{inputenc}
\usepackage[english]{babel}
\usepackage{academicons}
\definecolor{orcidlogocol}{HTML}{A6CE39}

\usepackage[pagewise]{lineno}
\nolinenumbers

\yil{2019}
\vol{??}
\fpage{}
\lpage{}
\doi{...}

\newcommand{\orch}{\includegraphics{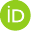}}
\usepackage{capt-of}

\title{CCD $UBVRI$ photometry of the open cluster Berkeley 8}

\author[\c{C}akmak et al.]{
  \textbf{
    Hikmet \c{C}AKMAK$^{1}$\thanks{hcakmak@istanbul.edu.tr}~\href{https://orcid.org/0000-0002-1959-6049}{\orch}, Ra\'ul MICHEL$^2$~\href{https://orcid.org/0000-0003-1263-808X}{\orch} and Y\"{u}ksel KARATA\c{S}$^{1}$~\href{https://orcid.org/0000-0002-7737-6589}{\orch}
   } \\
$^{1}$Department of Astronomy and Space Sciences, Faculty of Science, \.{I}stanbul University, 34116, \.{I}stanbul, Turkey\\
$^{2}$Observatorio Astron\'omico Nacional, Universidad Nacional Aut\'onoma de M\'exico, Apartado Postal 877,\\ 
 C.P. 22800, Ensenada, B.C., M\'exico\\
\\ [1.8em]

\rec{??.??.2019}
\acc{??.??.2019}
\finv{??.??.2019}
}

\newcommand{\aap}   {Astron. Astrophys. }
\newcommand{\aaps}  {Astron. Astrophys. Suppl. }

\newcommand{\aj}    {Astron. J. }

\newcommand{\mnras} {Mon. Not. Roy. Astron. Soc. }

\newcommand{\pasp}  {Pub. Astron. Soc. Pac. }
\newcommand{\pasj}  {Pub. Astron. Soc. Japan }

\newcommand{\astn}  {Astron. Nachr. } 
 
\newcommand{\rmxaa} {Rev. Mex. Astron. Astrofis } 

\newcommand{\araa}  {Annu. Rev. Astron. Astrophys. }

\newcommand{\etal}  {et al. }

\newcommand{\bc}{\begin{center}}
\newcommand{\ec}{\end{center}}

\renewcommand{\thefootnote}{\fnsymbol{footnote}}
\numberwithin{equation}{section}

\renewcommand{\phi}{\varphi}

\begin{document}
\maketitle

\begin{abstract}
The poorly studied Berkeley 8 (Be~8) open cluster is analysed from CCD $UBVRI$ photometric data taken with the 0.90~m telescope at the Sierra Nevada Observatory. The $Z = +0.008$ PARSEC isochrone gave us a reddening of E(B-V) = 0.69 $\pm$ 0.03, a distance of 3410 $\pm$ 300 pc and an age of 2.8 $\pm$ 0.2 Gyr. Its median Gaia DR2 distance, d~$=$~3676 $\pm$ 810 pc is in good agreement with our photometric distances, 3410 - 3620 pc within the uncertainties. The kinematic parameters of five likely members of Be~8 with the circular orbits, ecc $ = [0.23, ~0.30]$  reflect the properties of the Galactic thin disc, which is also consistent with what is expected of its metal content ([M/H] $=$ -0.27). Be 8 with R~$>$~9 kpc (co-rotation gap at 9 kpc) may have been originating from different galactic radius or different star formation region.

\end{abstract}

\renewcommand{\thefootnote}{\arabic{footnote}}
\setcounter{footnote}{-1} 

\section{Introduction}
Open clusters (OCs) are valuable objects for revealing stellar evolution and the structure, chemical and dynamical evolution of the Galactic disk. Be~8 has been studied by Hasegawa \etal (2004) \cite{has04} from CCD BVI photometry and by Bukowiecki \etal (2011) \cite{buk11} from 2MASS JHK$_{s}$ photometry.  Be~8 is located close to a portion of the Perseus spiral arm in the second quadrant of the Galaxy (Figure 1), according to its equatorial and Galactic coordinates (WEBDA) \cite{mer92} (rows 1--4 of Table 1). The main aim of this paper is to present astrophysical parameters such as reddening, distance and age of Be~8 from four colour indices, $(B-V)$, $(V-I)$, $(R-I)$ and $(G_{BP}$--$G_{RP})$ obtained from deep CCD $UBVRI$ and Gaia photometries. This kind of data is also valuable for classifying early-type stars, Blue Stragglers (BS) and Red Giant/Red Clump (RG/RC) candidates in the colour magnitude diagrams (CMDs), and thus probable  candidates are proposed for future spectroscopic observations. We used Gaia DR2 astrometric data (proper motion components and parallaxes) \cite{lin18,bro18} and Gaia DR2 photometry (G - G$_{BP}$G$_{RP}$) for determining the probable members of Be~8.  With the Gaia DR2 astrometric data, a membership method was done in the literature \cite{bal98,sar12,dia18}. The membership determinations of previous works have been based on the proper motions of Roeser \etal (2010) \cite{roe10} in combination with the 2MASS JHK$_{s}$ photometry of Skrutskie \etal (2006) \cite{skr06}. Cantat-Gaudin \etal (2018) \cite{can18} state that the proper motion uncertainties of UCAC4 fall in the range of 1--10 mas~yr$^{-1}$ \cite{roe10,zac13}. According to Lindegren \etal (2018) \cite{lin18} and Brown \etal (2018) \cite{bro18}, the mean parallax errors of Gaia DR2 catalogue fall in the range 0.02$–$0.04 mas  for G $<$ 15 and  0.1 mas for G $<$ 17, whereas the uncertainties of proper motion components are up to 0.06 mas~yr$^{-1}$ for G $<$ 15 mag and 0.2 mas~yr$^{-1}$ for G $<$ 17.   

This paper is organised as follows: Section~2 describes the observation and reduction techniques. Its dimensions are given in Section 3. Section 4 is devoted for the classification of cluster members. Section~5 describes the derivation of the astrophysical parameters. The classification of BS and RG/RC candidates and its morphological age determination are presented in Section 6. Section 7 focuses into its kinematics and orbital parameters. Discussions and Conclusions are given in Section~8.

\begin{figure}\label{Fig-1}
	\centering{\includegraphics[width=0.55\columnwidth]{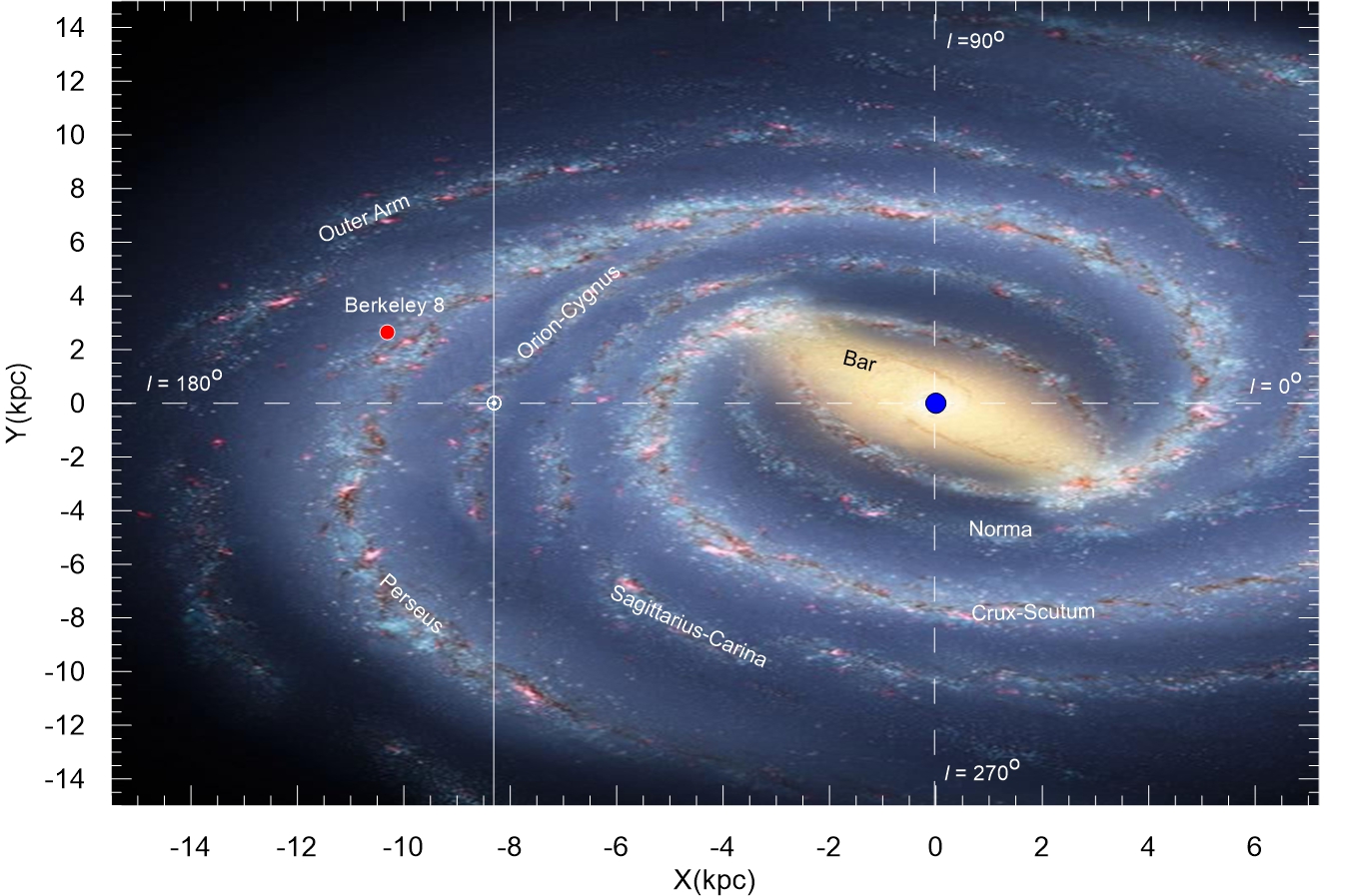}}
	\caption {Spatial distribution (X,~Y) (kpc) (filled red circle) of Be~8.  See Section 6 for the estimation of (X,~Y) values. The sun is at  R$_{\odot}$ = 8.2~$\pm$~0.1 kpc \cite{bg16}. The image is adapted from the image\protect\footnotemark credit by Robert Hurt, IPAC; Bill Saxton, NRAO/AUI/NSF.}
	\vspace*{-1.1em}
\end{figure}
\footnotetext{https://www.universetoday.com/102616/our-place-in-the-galactic-neighborhood-just-got-an-upgrade}

\renewcommand{\arraystretch}{1.3} 
\begin{table}[!t]\label{Table-1}
	\centering
		\caption{Rows~(1)-(4) mean the central equatorial (J2000) and the Galactic coordinates of WEBDA. Air mass and exposure times (s) of filters are listed in rows 5--10.}
		\label{tab:coordinates-PM}
		\setlength{\tabcolsep}{1.8\tabcolsep}
			\begin{tabular}{lr}
		\hline
			Cluster  & Be~8  \\
			\hline
			$\alpha_{2000}$ (h\,m\,s) & 02 01 08.00 \\
			$\delta_{2000}$ $(^{\circ}\,^{\prime}\,^{\prime\prime})$ & $+$75 29 38.25 \\
			$\ell$ $(^{\circ})$ & 127.35  \\
			b $(^{\circ})$ & $+$13.21 \\
			Air mass & 1.276 -- 1.280 \\
			Filter U Exp.Time (s) & 9, 90, 3x300 \\
			Filter B Exp.Time (s) & 5, 50, 500  \\
			Filter V Exp.Time (s) & 3, 30, 300  \\
			Filter R Exp.Time (s) & 2, 20, 200 \\
			Filter I \, Exp.Time (s)&2, 20, 200 \\
			\hline
		\end{tabular}
	\end{table}

\section{Observations and Reduction techniques}

Observations of Be~8 were carried out, during the photometric night of December 3 2015, with the 0.90m (f/8 Ritchey-Chrétien) telescope at Sierra Nevada Observatory (Granada, Spain). A filter wheel with $UBVRI$ filters and a scientific grade Marconi-EEV CCD42-40 were employed. The CCD is a 2048x2048 13.5-$\mu$m square-pixel  detector with a nominal gain of 1.35 e$^-$/ADU and a readout noise of 7.14 e$^-$ at the 2$\times$2 binning employed. Along with the optics, it covers a field of view of 13.2$\times$13.2 arcmin$^2$. Apart form Be~8, other open clusters and some Landolt standard fields \cite{lan09} were observed. Flat fields were also acquired at the beginning of the night and many bias frames were also taken.

\begin{figure}[!t]\label{Fig-2}
	\begin {center}
	\includegraphics[width=0.4\columnwidth]{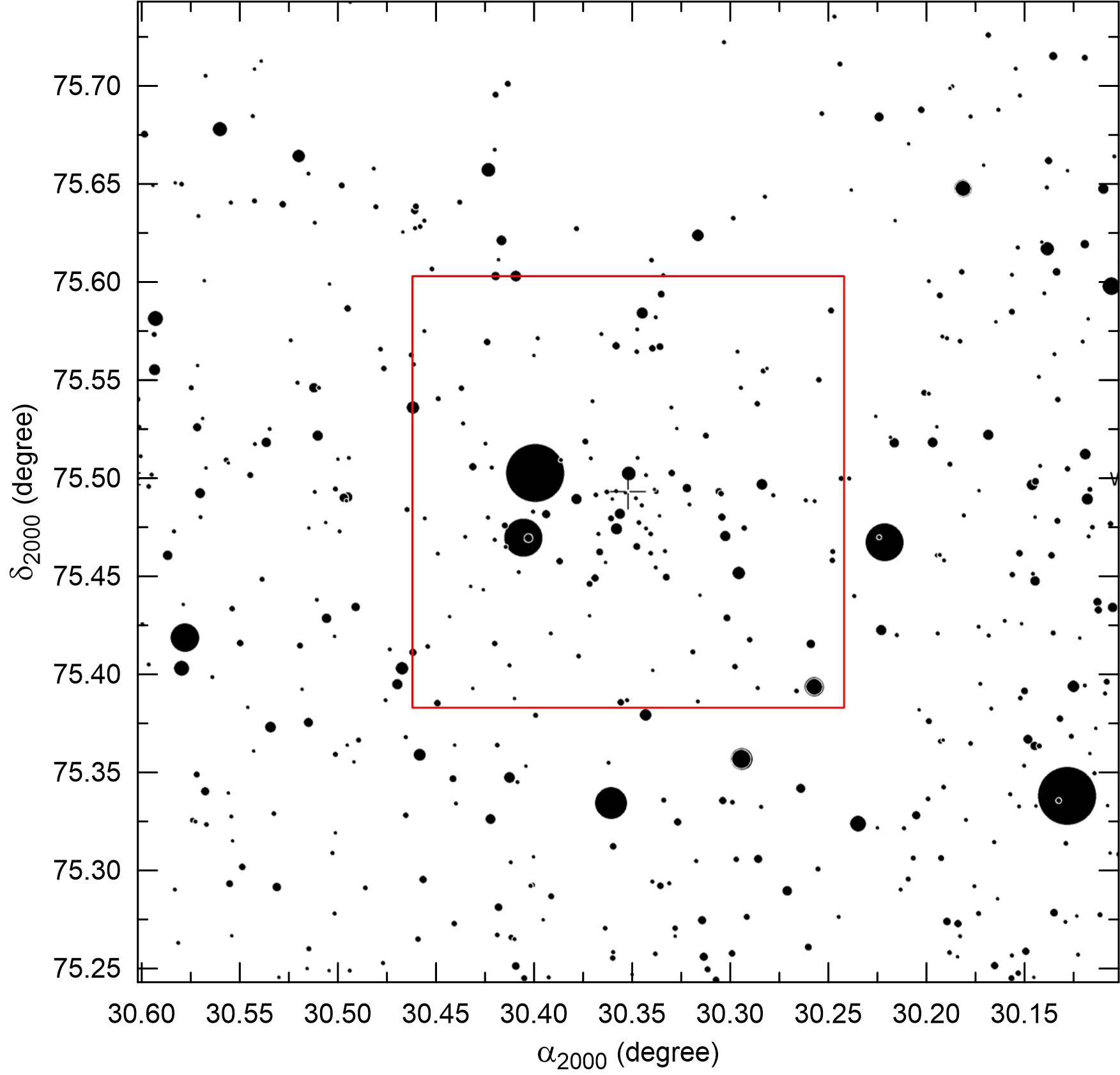}
	\caption{The image of Be~8 for a region of $18.3\,^\prime$x$17.2\,^\prime$ (https://www.aavso.org (AAVSO). The
		red rectangle indicates the field of view of the SNO detector, $13.2^{\prime}\times13.2^{\prime}$.}
	\end {center}
\end{figure}

The data reduction was carried out by R. Michel \footnote {Data may be requested from R. Michel.} using the IRAF\footnote {IRAF is distributed by the National Optical Observatories, operated by the Association of Universities for Research in Astronomy,	Inc., under  cooperative agreement with the National Science Foundation.} package and together with some home-made auxiliary Fortran programs and Awk scripts. All the images were bias subtracted and flat-field corrected (CCDRED). Cosmic rays were removed with the L.A. Cosmic\footnote{http://www.astro.yale.edu/dokkum/lacosmic} script van Dokkum (2001) \cite{van01}. 

The standard magnitudes were taken from the catalogue by Landolt (2009) \cite{lan09} and supplemented with the secondary photometric standards by Cutri \etal (2013) \cite{cut13}. As a result, the transformation coefficients were found (FITPARAMS). For magnitude estimation, transformation equations used are

\begin{flalign}
M_{\lambda} &=  m_{\lambda}-(k_{1}-k_{2}C)X + \eta_{\lambda} C +\zeta_{\lambda}
\end{flalign}
\noindent
where $m_{\lambda}$, $k_{1}$, $k_{2}$, $C$ and $X$ are observed instrumental magnitude, primary and secondary extinction coefficients, colour index and air mass, respectively. $M_{\lambda}$, $\eta_{\lambda}$, $\zeta_{\lambda}$ are standard magnitude, atmospheric extinction-corrected instrumental magnitude, transformation coefficient and photometric zero point, respectively. The other details of data reduction can be found in the papers of Akkaya \etal (2010) \cite{akk10} and Akkaya \etal (2015) \cite{akk15}. Air-mass range and exposure times in each band during the observations are given in rows 5--10 of Table 1.	
 
\begin{figure*}[!t]\label{Fig-3}
	\centerline{\includegraphics[width=0.35\columnwidth]{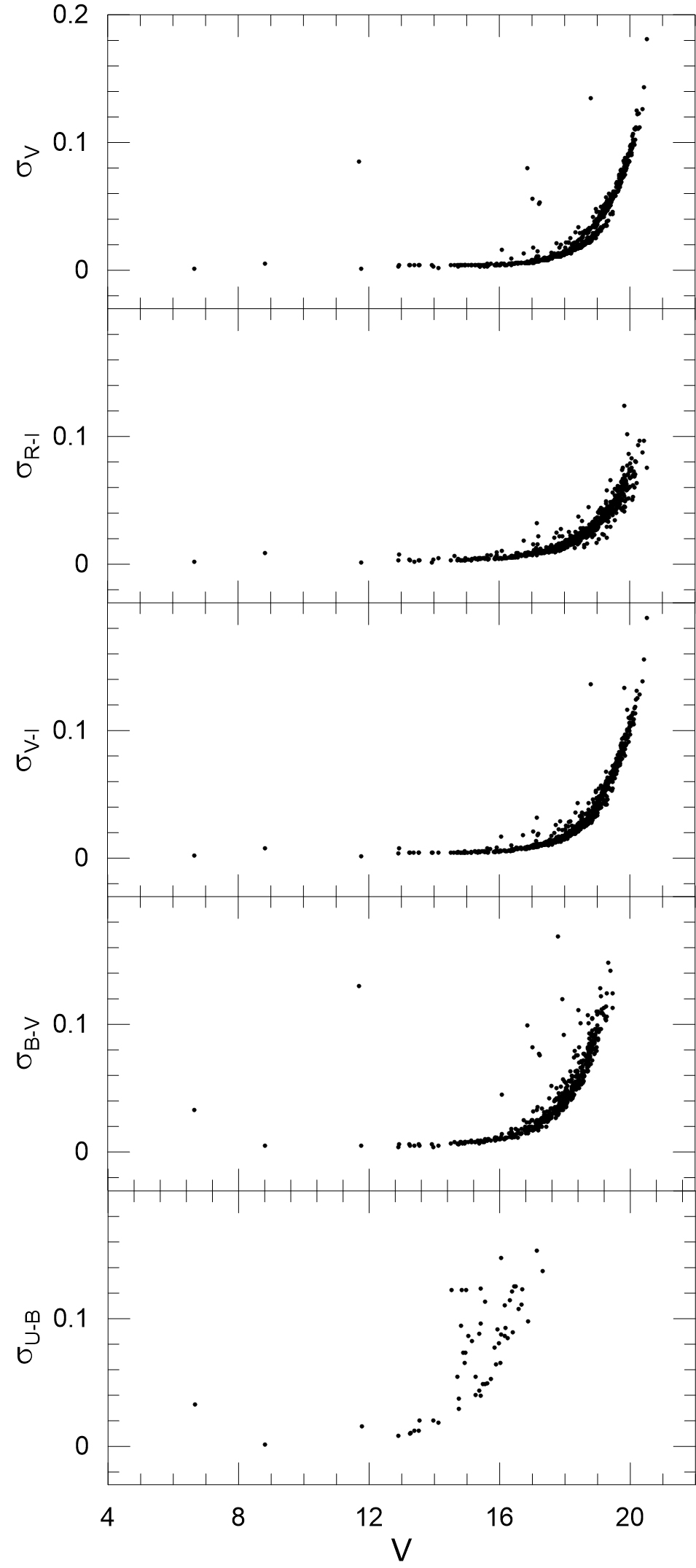}}
	\caption{Photometric errors of the $V$ apparent magnitude and four colours against the $V$ magnitude for Be~8.}
\end{figure*}

\renewcommand{\arraystretch}{1.2}
\begin{table}[!h]\label{Table-2}
	\caption {The mean photometric errors of $V$, $(U-B)$, $(B-V)$, $(R-I)$, $(V-I)$ of Be~8.}
	\setlength{\tabcolsep}{1.3\tabcolsep}  \vspace{2ex} \centering {
		\begin{tabular}{ccrrrrr}
			\hline
			V &$\sigma_{V}$&$\sigma_{U-B}$&$\sigma_{B-V}$ &$\sigma_{R-I}$ &$\sigma_{V-I}$&N \\
			\hline
			6 - 13  &  0.017 & 0.014 & 0.031 &  0.004 &  0.004 &  6 \\
			13 - 14 &  0.004 & 0.014 & 0.005 &  0.003 &  0.004 &  7 \\
			14 - 15 &  0.004 & 0.074 & 0.007 &  0.004 &  0.004 & 13 \\
			15 - 16 &  0.004 & 0.071 & 0.009 &  0.005 &  0.005 & 30 \\
			16 - 17 &  0.007 & 0.104 & 0.018 &  0.007 &  0.009 & 50 \\
			17 - 18 &  0.011 & 0.145 & 0.034 &  0.012 &  0.015 &166 \\
			18 - 19 &  0.022 &     - & 0.063 &  0.023 &  0.027 &265 \\
			19 - 20 &  0.058 &     - & 0.110 &  0.047 &  0.066 &216 \\
			20 - 21 & 0.107  &     - &     - &  0.073 &  0.117 & 30 \\
			\hline
	\end{tabular}  }
\end{table}

Figure 2 presents the finding chart \footnote{Obtained from \textit{https://www.aavso.org (AAVSO)} web page, [accessed 09 June 2019]} of Be~8 ($18.3\,^\prime$x $17.2\,^\prime$). The red rectangle indicates the field of view of the SNO detector, $13.2^{\prime}\times13.2^{\prime}$.
The photometric errors in $V$ and $(R-I)$, $(V-I)$, $(B-V)$, $(U-B)$ of Be~8 are presented in Figure 3. Its  mean photometric errors are also listed in Table 2. Our inspection of Figure 3 and Table~2 indicates that stars brighter than $V = 18$ have errors smaller than 0.03 in $(R-I)$, $(V-I)$, $(B-V)$. For $V > 20$,  the errors in $(R-I)$ and $(V-I)$ are larger than 0.03.  After $V > 18$,  the errors in $(B-V)$ are up to $\approx0^m.03$.  The errors in $(U-B)$ are less than 0.01 for $V < 14$, whereas for $14 < V \le 18$, large errors increase.

\section{Dimensions of Be~8}

The stellar radial density profile (RDP) of Be~8 (Figure 4) has been built from Gaia DR2 photometry for the equatorial coordinates (Table~1). Its RDP have been constructed by counting stars in concentric rings of increasing width with distance to its centre. We choose $\Delta R= 15.0^{\prime}$ as the wide external ring of the stellar comparison field. As emphasised by Bonatto and Bica (2007) \cite{bon07}, the number and width of rings were optimised so that the resulting RDP had adequate spatial resolution with moderate 1$\sigma$ Poisson errors. The solid curve (Figure 4) denotes the fitted King's profile \cite{kin66}. Here we adopt the two-parameter function, $\sigma(R) =  \sigma_{bg} + \sigma_0/(1+(R/R_c)^2)$, where $\sigma_{bg}$  is the residual background density, $\sigma_0$ the central  density of stars, and R$_{core}$ the core radius. 
The horizontal red bar shows the stellar background level measured in the comparison field, and the $1\sigma$ profile  fit uncertainty is shown by the shaded region. 
The core and cluster radii from Figure 4 have been determined as $R_{core} =1.8^{\prime}$ and $R_{RDP}~=~15.0^{\prime}$, respectively. These dimensions are quite close to the ones of $(R_{core},~R_{RDP})=(1.52^{\prime}, ~15.5^{\prime})$ of  Bukowiecki \etal (2011) \cite{buk11}.

\begin{figure}[!h]\label{Fig-4}
	\begin {center}
	\includegraphics[width=0.40\columnwidth]{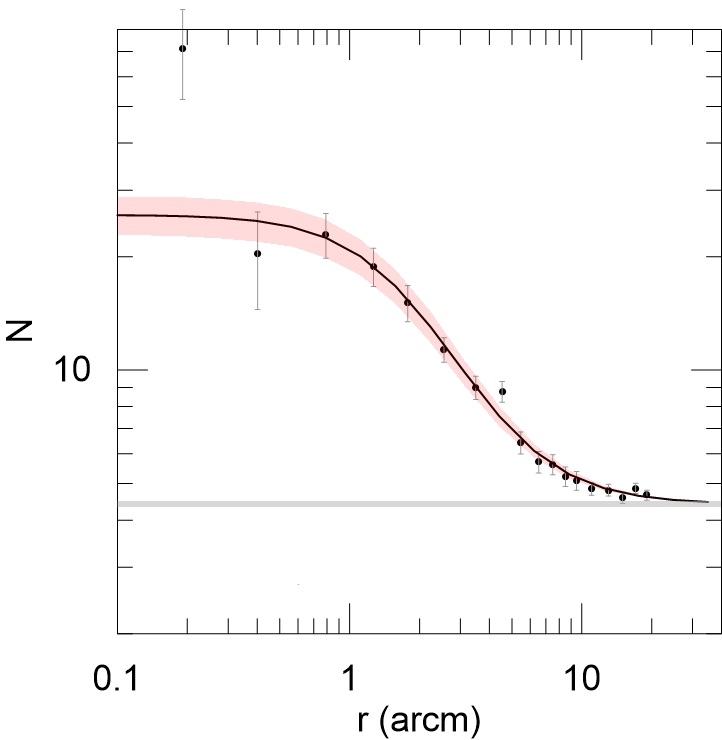}
	\caption{Stellar RDP (filled dots) of Be~8. Solid line shows the best-fit King profile. Horizontal red bar: stellar background level measured  in the comparison field. Shaded region: $1\sigma$ King fit uncertainty. The core and cluster radii are obtained as $R_{core} = 1.8^{\prime}$ and $R_{RDP} = 15.0^{\prime}$, respectively.}
	\end {center}
\end{figure}

\section{Classification of cluster members}

For the membership determination of Be~8, our CCD $UBVRI$ photometric data have been matched with Gaia DR2 astrometric  (proper motion components and parallaxes) ($\mu_{\alpha}$, $\mu_{\delta}$ and $\varpi$) and Gaia DR2 photometric data. The $\mu_{\alpha}$ versus $\mu_{\delta}$ ($mas~yr^{-1}$) for all stars of Be~8 (filled dots) is shown in Figure 5. Grey dots denote the field stars inside $R=15'.0$ arcmin centred on Be~8. We have applied Gaussian Mixture method for all stars of Be~8.

By applying Gaussian Mixture Model (GMM) \cite{ped11} to the stars in the cluster region of Be~8, we have determined the membership probability ($P$). The model considers that the distribution of proper motions of the stars in a cluster's region can be represented by two elliptical bivariate Gaussians, by following Wu \etal (2002) \cite{wu02} which include the proper motion's errors in the frequency function. The expressions used can be found in the papers of Balaquer-Nunez \etal \cite{bal98}, Sariya \etal (2012) \cite{sar12} and Dias \etal (2018) \cite{dia18}.
$P$ is defined $\Phi_c$ /$\Phi$.  Here $\Phi = \Phi_c + \Phi_f$ is the total probability distribution. \textit{c} and \textit{f} subscripts for cluster and field parameters, respectively. The used parameters for estimation of $\Phi_c$ and $\Phi_f$ are $\mu_{\alpha}$, $\mu_{\delta}$, $\varpi$, $\sigma_{\mu\alpha}$, $\sigma_{\mu\delta}$, $\sigma_\varpi$. 
The averages and their uncertainties of proper motion components and parallaxes in  the distributions of the cluster and field regions in Figure 5 are listed in Table~3. Figure 6(a) shows the membership probability ($P$) histogram, which provides a very clear separation between cluster and field stars. The number of stars with membership probability which is greater than 90$\%$ is 273. These likely members have been considered for deriving astrophysical parameters of Be~8. From the relation of the membership probability ($P$) versus G mag (Figure 6(b)), high membership probability appears to extend down to G $\sim$ 21 mag. 

\begin{figure*}[!t]\label{Fig-5}
	\begin {center}
	\includegraphics[width=0.45\columnwidth]{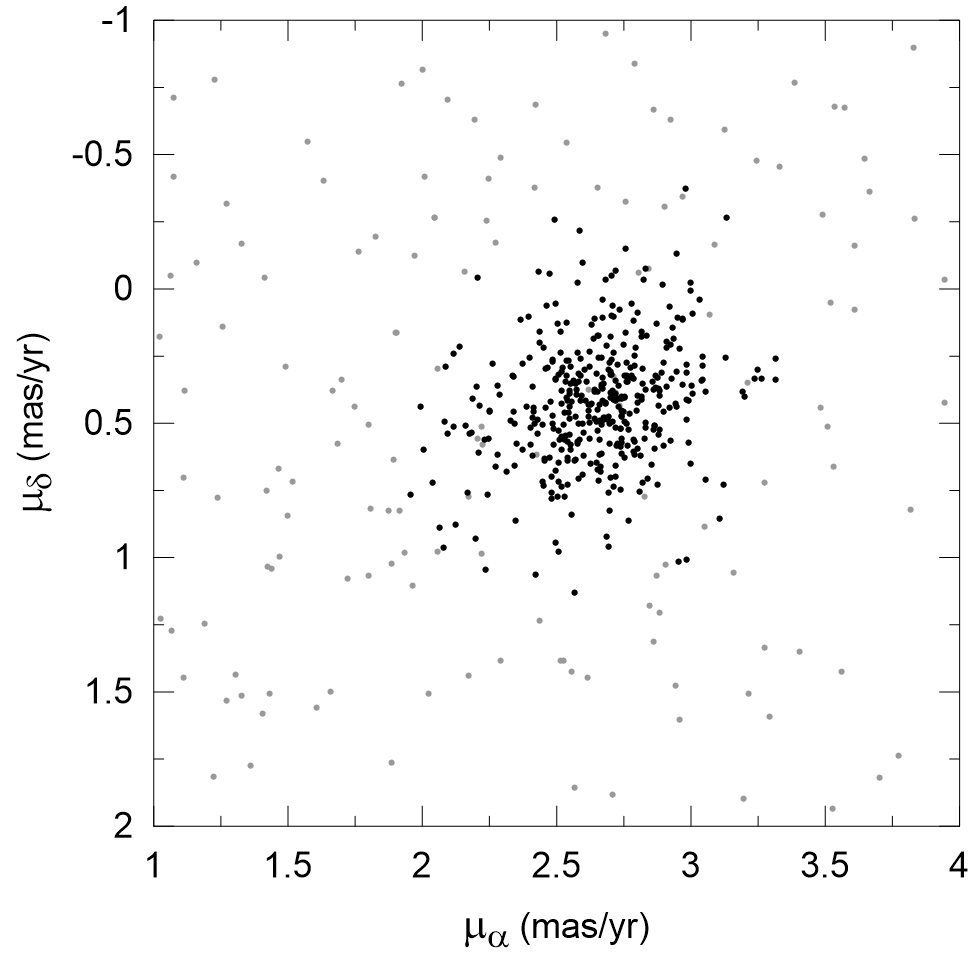}
	\caption{The $\mu_{\alpha}$ versus $\mu_{\delta}$ for Be~8 (filled stars). Small grey dots represent the Gaia DR2 astrometric data for a  $15'.0$ field centred on Be~8.}
	\end {center}
\end{figure*}

\renewcommand{\arraystretch}{1.5}
\begin{table}[!h]\label{Table-3}
	\caption {The medians and their uncertainties of proper motion components  ($\mu_{\alpha}$,~ $\mu_{\delta}$) ($mas~yr^{-1}$) and parallaxes ($\varpi$) (mas) from the distributions of the cluster and field regions on Figure 5.}
	\setlength{\tabcolsep}{2.0\tabcolsep}  \vspace{2ex} \centering {
		\begin{tabular}{rrrr}
			\hline
			&$\mu_{\alpha}$ $\pm$ $\sigma_{\mu\alpha}$ & $\mu_{\delta}$ $\pm$ $\sigma_{\mu\delta}$ &$\varpi$ $\pm$ $\sigma_\varpi$ \\
			\hline
			Cluster region & 2.648 $\pm$  0.24 & 0.421 $\pm$ 0.23 & 0.266 $\pm$ 0.12 \\
			Field region   & 0.33 $\pm$ 2.23 & 0.347 $\pm$ 1.75 & 0.545 $\pm$  0.45 \\
			\hline
	\end{tabular}  }
\end{table}

\begin{figure}[!t]\label{Fig-6}
	\begin {center}
	\includegraphics[width=0.42\columnwidth]{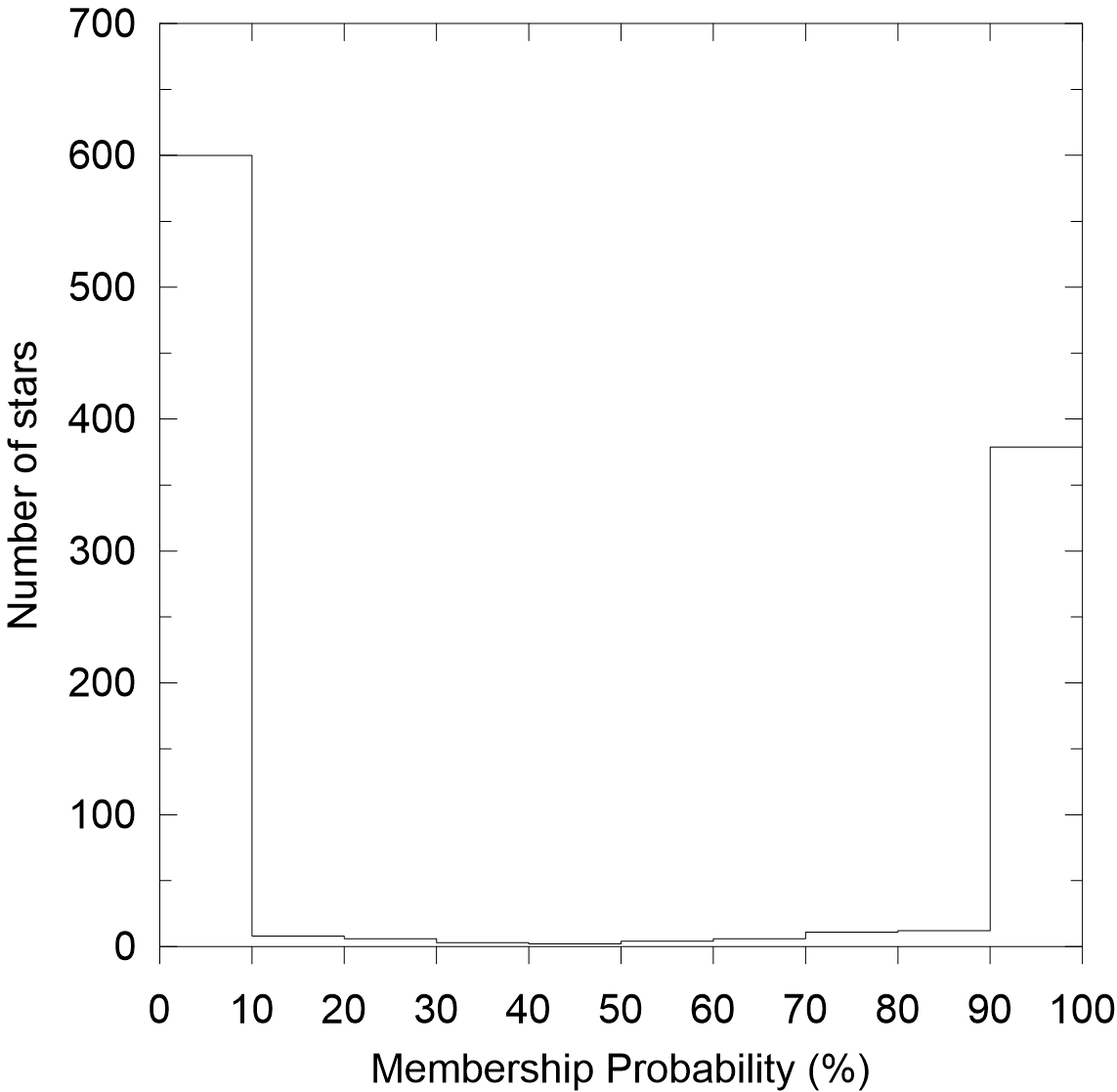}\hspace{0.8cm}
	\includegraphics[width=0.42\columnwidth]{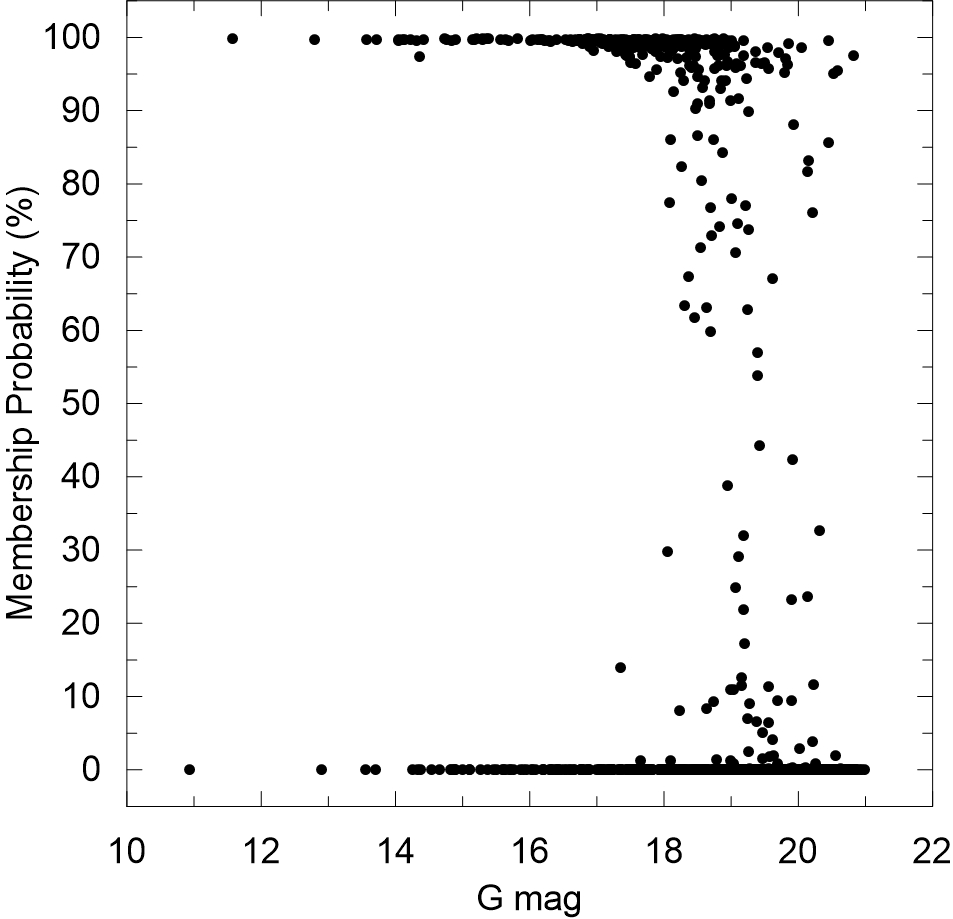}
	\caption{The membership probability histogram P($\%$) (left panel) and P($\%$) versus G mag (right panel) for all stars of Be~8.}
	\end {center}
\end{figure}

\begin{figure}[!h]\label{Fig-7}
	\begin {center}
	\includegraphics[width=0.4\columnwidth]{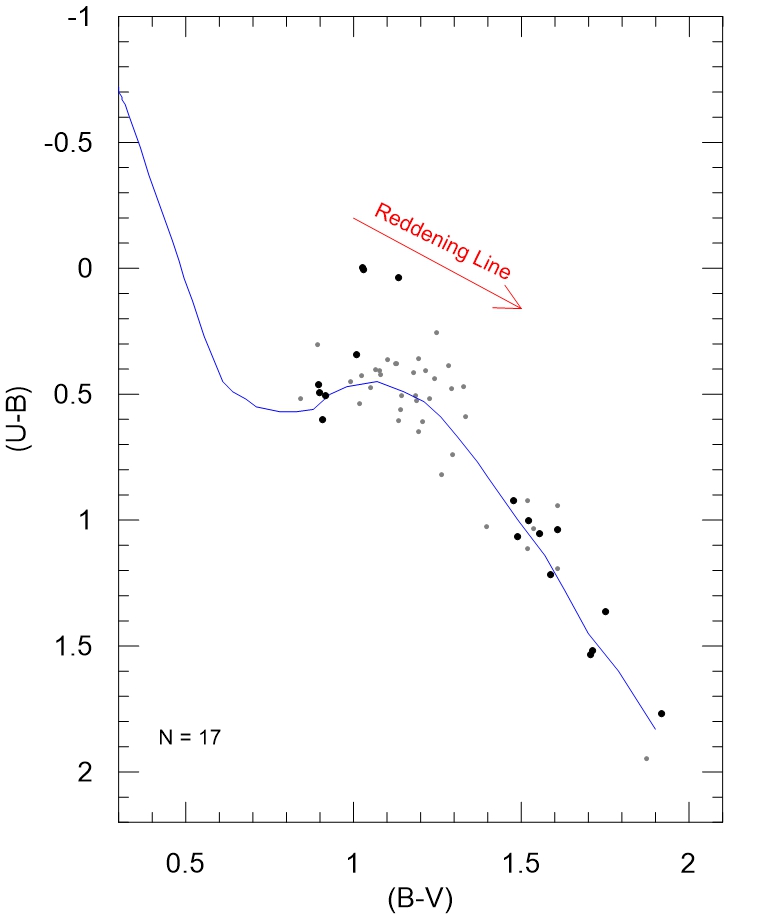}
	\caption{$(U-B)$ - $(B-V)$ two colour diagram for the 17 likely members (filled dots) of Be~8. The dashed blue line represents the Schmidt-Kaler (SK82) main-sequence. Field stars are shown with small grey dots. Red arrow denotes the reddening.}
	\end {center}
\end{figure}

\section{Astrophysical parameters of Be~8 open cluster}

The two colour $(U-B)$ - $(B-V)$ (CC) diagram of the 17 probable members for Be~8 is displayed in Figure 7.  
The blue line denotes the reddened main sequence for dwarfs and red giants of Schmidt-Kaler (1982)(SK82) \cite{sch82}. It appears from Fig. 7 that Be~8 contains F-type stars which are quite valuable for determining photometric metal abundance, [M/H]. These F type stars with $(U-B) < $ 0.35 and $0.9 < (B-V) < 1.2$ have large UV-excesses, $\delta (U-B) = 0.15-0.48$. A member star with $\delta (U-B) = 0.15$ estimates $[M/H] = -1.0$, which is very poor for Galactic open clusters. The other three members with $\delta (U-B) = 0.47 - 0.48$ do not allow us determine a reasonable [M/H] value, due to their large UV-excesses which are insensitive to the metal abundance calibration (see the paper of Karata\c{s} and Schuster (2006) \cite{kar06}). Moreover, their photometric errors in $(U-B)$ are larger at a level of $\sigma_{(U-B)} = 0.10 - 0.15$.  

\renewcommand{\arraystretch}{1.4} 
\begin{table}[!t]\label{Table-4}
	\begin {center}
	\caption{The derived fundamental astrophysical parameters of Be~8 for four colour indices.}
	\label{tab:be8parameters} 
	\setlength{\tabcolsep}{0.5cm}
	\begin{tabular}{lcccc}
		\hline
		Colour &$(V_{0}$--$M_{V})$ &  d~(kpc) & log(A)  & A~(Gyr) \\
		\hline
		$(B-V)$ &12.66 $\pm$ 0.19 &3.41 $\pm$ 0.30 &9.45 $\pm$ 0.03 &2.80 $\pm$ 0.20 \\
		$(V-I)$ &12.67 $\pm$ 0.31 &3.42 $\pm$ 0.49 &9.45 $\pm$ 0.03 &2.80 $\pm$ 0.20 \\
		$(R-I)$ &12.79 $\pm$ 0.32 &3.62 $\pm$ 0.53 &9.45 $\pm$ 0.03 &2.80 $\pm$ 0.20 \\
		$(G_{BP}-G_{RP})$ &12.71 $\pm$ 0.31 &3.48 $\pm$ 0.50 &9.45 $\pm$ 0.03 &2.80 $\pm$ 0.20 \\
		\hline
	\end{tabular} 
	\end {center}
\end{table}   

\begin{figure}[!b]\label{Fig-8}
	\begin {center}
	\includegraphics[width=0.4\columnwidth]{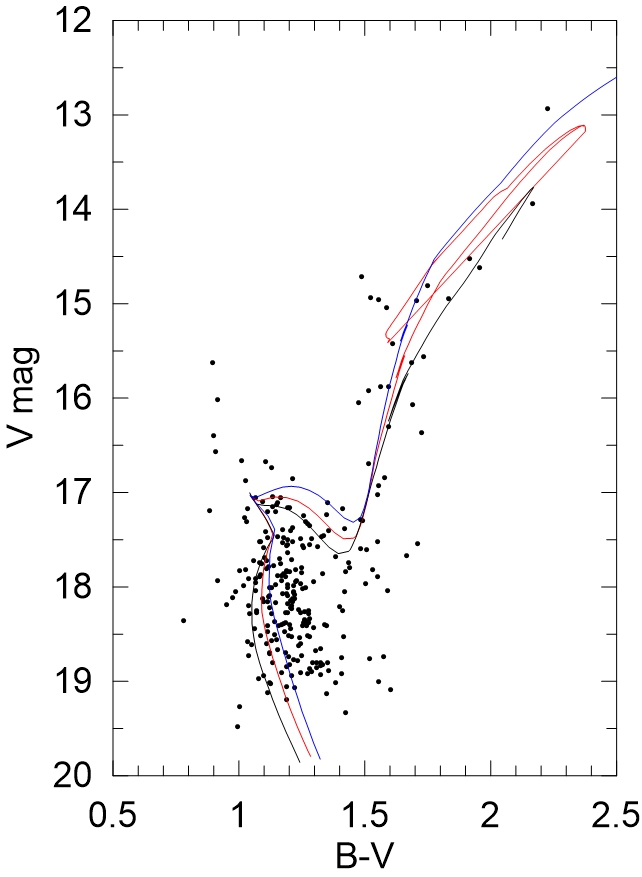}\hspace{1.2cm}
	\includegraphics[width=0.39\columnwidth]{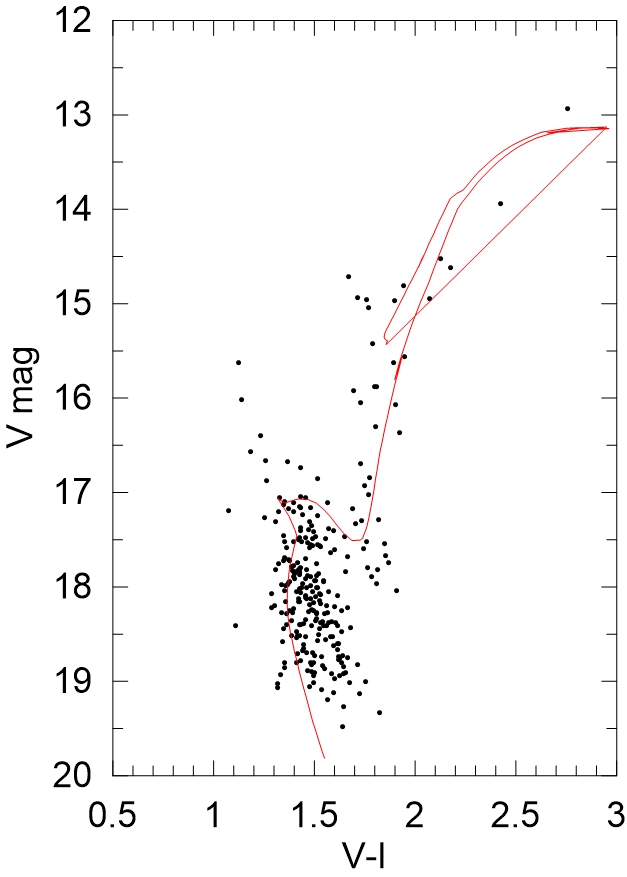}
	\caption{For the 268 likely members of Be~8, CMDs of $V$ - $(B-V)$ (\textit{Left}) and $V$ - $(V-I)$ (\textit{Right}). Red curves show the PARSEC isochrones interpolated to $Z = +0.008$. In \textit{Left} figure the fitted isochrones $Z = +0.015$ (blue line) and +0.004 (black line) are also plotted (See Section 5).}
	\end {center}
\end{figure}

\begin{figure}[!t]\label{Fig-9}
	\begin {center}
	\includegraphics[width=0.41\columnwidth]{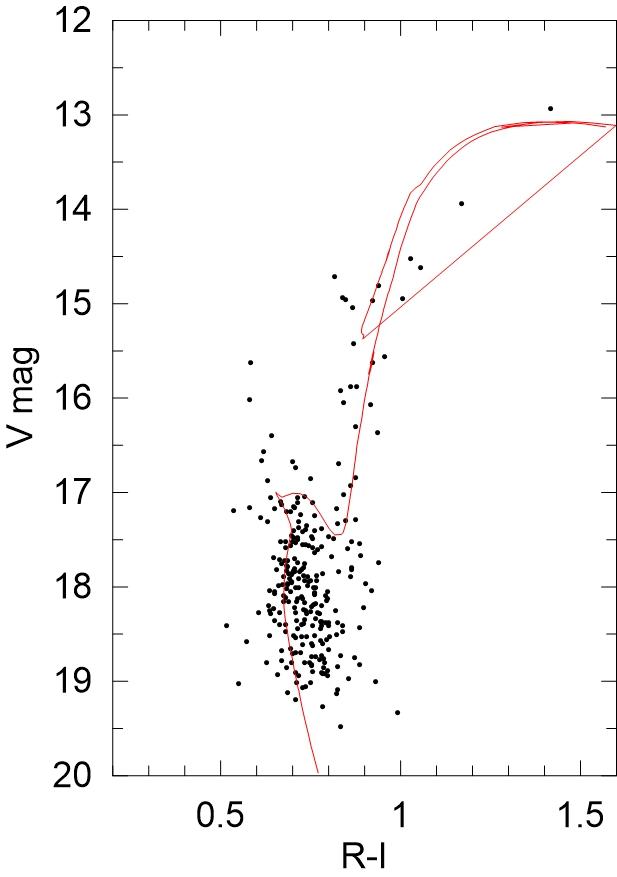}\hspace{1.2cm}
	\includegraphics[width=0.41\columnwidth]{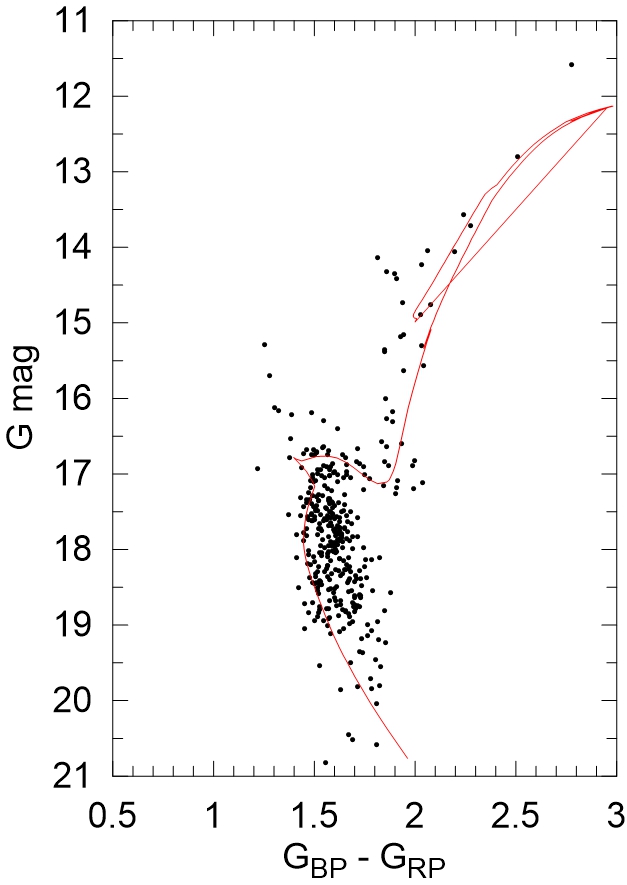}
	\caption{For the 268 likely members of Be~8, CMDs of $V$ - $(R-I)$ (\textit{Left}) and $G$ - ($G_{BP}-G_{RP}$) (\textit{Right}). The meanings of the symbols are the same as Figure 8.}
	\end {center}
\end{figure}

Since these issues make it difficult to determine its photometric metal abundance and reddening from CC diagram, we have derived the reddenings, distance moduli and ages of Be~8 from the PARSEC isochrones of Bressan \etal (2012) \cite{bre12} on the CMDs for four colour indices. The appropriate PARSEC isochrones for different heavy element abundance mass fractions (Z $=$ +0.015, +0.004, +0.008) (left panel of Figure 8) and reddenings have been fitted on the CMDs. The 2.8 Gyr PARSEC isochrones for Z $=$ +0.008 abundance gave us a good fit solution on the CMDs: $V$ - $(B-V)$, $V$ - $(V-I)$, $V$ - $(R-I)$, $G$ - ($G_{BP}-G_{RP}$) (Figs.~8--9). The equation $Z = Z_\odot 10^{[M/H]}$ estimates its photometric metal abundance as [M/H] $=-$0.27. Here  the solar heavy metal content is adopted as $Z_{\odot} = +0.015$. The isochrone is varied until a satisfactory fit to the data has been obtained through the observed main-sequence (MS), turn-off (TO), sub-giant (SG) and Red Giant/Red Clump (RG/RC) sequences on the CMDs, as we followed in the papers of Akkaya \etal (2010) \cite{akk10}, Akkaya \etal (2015) \cite{akk15} and G\"{u}ne\c{s} \etal (2012) \cite{gun12}.  Because of the presence of binaries, the PARSEC isochrones have been shifted to the left and below of MS on all CMDs. The vertical shift gives the (true) distance modulus, $DM = (V_{0}-M_{\rm V})$. For its age (A,~log(\rm A)), the PARSEC isochrones have been shifted both vertically and horizontally on the CMD's with the expression $M_{V}+3.1E(B-V)+DM$, for the vertical displacement and $C_0(\lambda_1 - \lambda_2)+E(\lambda_1 - \lambda_2)$, for the horizontal, where $\lambda$ denotes the wavelengths of $BVRI$ and $G$, $G_{BP}$, $G_{RP}$  filters. Here C$_{0}$ means de-reddened colour index. The obtained reddenings from four CMDs are E(B$-$V) = 0.69 $\pm$ 0.08, E(V$-$I) = 0.87 $\pm$ 0.10, E(R$-$I) = 0.44 $\pm$ 0.05,  E($G_{BP}-G_{RP}$) = 0.91 $\pm$ 0.10, respectively. Its distance moduli ($V_{0}$ -- $M_{\rm V}$)/ distances (d~(kpc)) and age (Gyr) for four colours are presented in Cols.~2--5 of Table~4. In order to  de-redden the distance moduli (Col.~2 of Table 4) for the colour indices $(V-I)$, $(R-I)$, ($G_{BP}-G_{RP}$), our colour excesses E(V~$-$~I), E(R$-$I) and E($G_{BP}-G_{RP}$) have been converted from the relations E(V~$-$~I) = 1.25 E(B$-$V), E(R$-$I) = 0.69 E(B$-$V) \cite{dea78, math90, str95} and E(B$-$V) = 0.775 E(G$_{BP}-G_{RP}$)\cite{bra18}. 

\section{Blue Stragglers, Red Giants/Clumps and Morphological age index}
From Figs.~8 -- 9 we note that six probable Blue Straggler (BS) candidates lie on $(B-V) < 1.1$, $(V-I) < 1.4$, $(R-I) < 0.7$ for $V < 17.0$, whereas they occupy the region with $G < 16.75$ and $G_{BP}$ - $G_{RP} < 1.4$. BS candidates blur the main-sequence turn off (TO) to a brighter magnitude. RG/RC candidates populate well the red-giant branch (Figs.~8 -- 9). These RG/RC stars in old clusters are numerous and luminous as expected. BS (blue plus symbol) and RG/RC (red plus) candidates with P $>$ 90$\%$ have been placed on the nested circles of Figure 10. Two BS candidates reside in core radius of 1.8~$^{\prime}$ of Be~8. Four BS stars occupy the inner regions. 26 RG/RC candidates remain within the cluster radius (R$_{RDP} = 15.0^{\prime}$). Their median Gaia DR2 distances are d $=$ 3953 $\pm$ 750 pc ($\varpi$~=~0.253 $\pm$ 0.048 mas) (six BS) and d $=$ 4049 $\pm$ 610 pc ($\varpi$~=~0.247 $\pm$ 0.037 mas)  (26 RG/RC), respectively, which are in reasonable agreement with the photometric ones within the uncertainties (Table 4). However, spectroscopic observations of these candidates are needed for their membership confirmation.

\begin{figure}[!b]\label{Fig-10}
	\begin {center}
	\includegraphics[width=0.5\columnwidth]{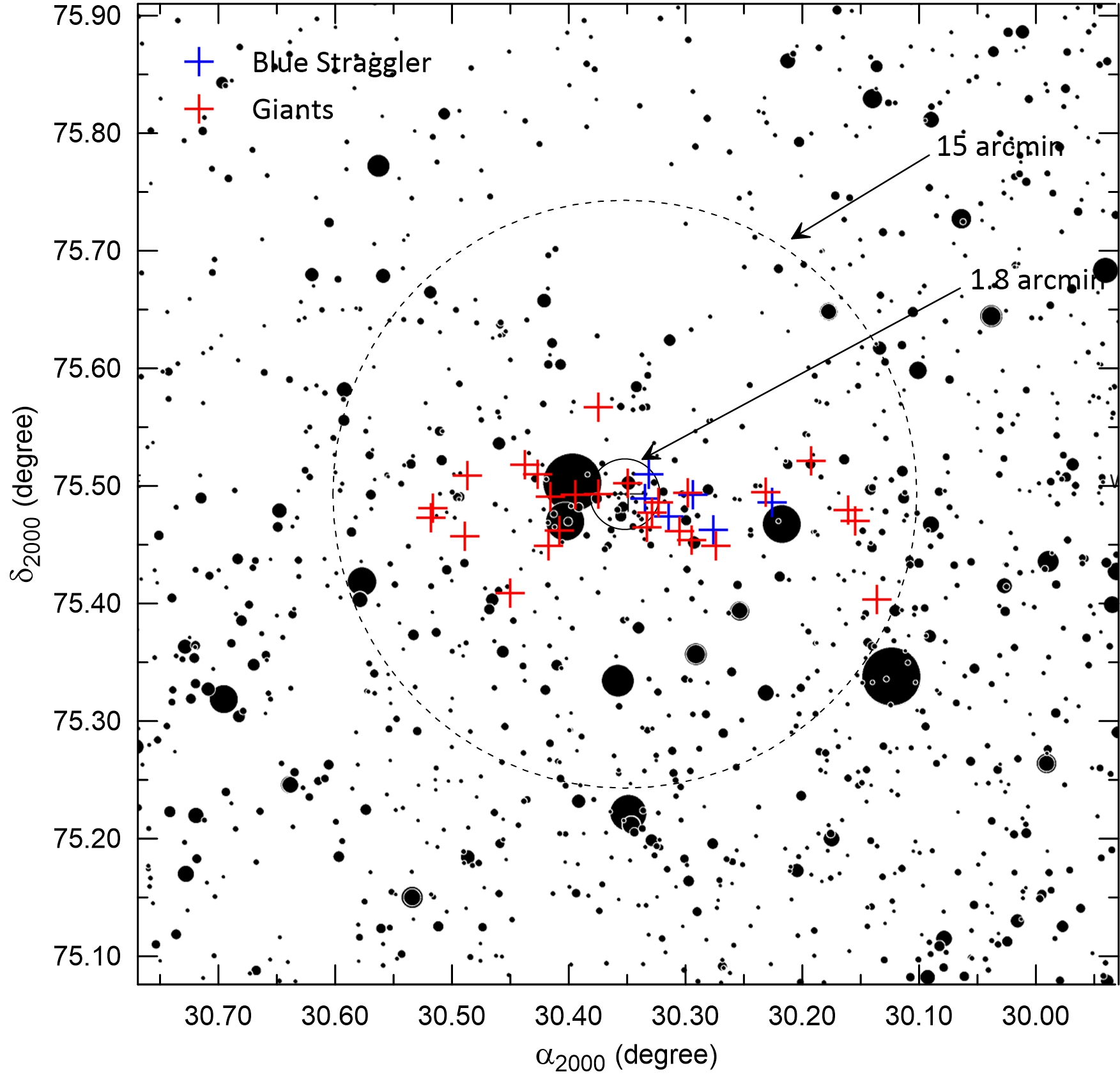}
	\caption{The circles have been drawn for the radii of $1.8^{\prime}$ and $15.0^{\prime}$ arc min on the $\alpha_{2000}$ versus $\delta_{2000}$ of Be~8. Red and blue pluses represent the RG/RC and BS candidates on the CMDs (Figs.~8 -- 9).}
	\end {center}
\end{figure}

The CMDs of Be~8 exhibit noticeable TO and RG/RC sequences (Figs.~8 -- 9). By utilising the definition of the morphological age index (MAI), $\delta 1 = (B $ - $ V)_{TO}-(B $ - $ V)_{RG}$ given by Phelps \etal (1994) \cite{ph94},  we measured V$_{TO}$ = 17.07, (B$-$V)$_{TO}$ = 1.066 and (B$-$V)$_{RG}$ = 1.599 on $V$ - $(B-V)$ diagram (left panel of Figure 8). These values estimate $\delta 1$ = 0.533. Here $\delta 1$ is the difference in the colour indices between the bluest point of TO and the colour at the base of RG branch one magnitude brighter than the TO luminosity. Our $\delta 1$ value has been transformed into $\delta V$ via the equation of $\delta V = 3.77 - 3.75 \delta 1$ of Phelps \etal (1994) \cite{ph94}. Its morphological age has been estimated as log A $=$ 9.45 (A $=$ 2.94 Gyr) from the equation, $log A = 0.04 \delta V^2+0.34 \delta V + 0.07[Fe/H] + 8.76$ of Salaris \etal (2004) \cite{sal04}, applying its the metal abundance ([M/H] $=-$0.27) and $\delta V$ = 1.771. Here we assume [Fe/H] $\approx$ [M/H]. Its MAI age (2.94 Gyr) is in good compatible with its isochrone age, 2.80 Gyr.

\section{Kinematics and Orbital parameters of Be~8}

Five likely members with Gaia DR2 radial velocities (Table 5) allow us to calculate heliocentric velocities, ($U$, $V$, $W$) from the algorithm of Johnson and Soderblom (1987) \cite{joh87}. These five members seem to be giant candidates according to their $V$ and $(B-V)$ values (left panel of Figure 8). The calculated $U$, $V$, $W$ velocities have been transformed to the components $U'$, $V'$, $W'$ by correcting for the Solar motion $(U, V, W)_{\odot} = (+11.10, +12.24, +7.25)$ km s$^{-1}$ with respect to the local standard of rest (LSR) \cite{sch10}. We adopt R$_{\odot}$$=$8.2$\pm$0.1 kpc \cite{bg16} and $V_{LSR}$ = 239\,km\,s$^{-1}$ \cite{bru11}. We adopt the right-hand system for the estimations. Their estimated heliocentric cartesian distances ($x'$, $y'$, $z'$) (kpc) and LRS-velocity components ($U'$, $V'$, $W'$) have been transformed to Galactic Rest of Frame (GSR) i.e., ($x$, $y$, $z$) (kpc) and ($V_{x}$, $V_{y}$, $V_{z}$) via the equations given by Kepley \etal (2007) \cite{kep07}. 
The Galactocentric velocity component ($V_{\Phi}$) (km s$^{-1}$) (or azimuthal velocity) in a cylindrical frame is estimated via $V_{\Phi} =  \frac{x V_{y} - y V_{x}}{R}$. $V_{\Phi}<0$ means prograde. Thus, the obtained  kinematic parameters (U, V, W, V$_{\Phi}$) km s$^{-1}$ are listed in Cols.~2 -- 5 of Table 6. 

\renewcommand{\arraystretch}{1.4} 
\begin{table*}[!t]\label{Table-5}
	{\small
		\begin {center}
		\caption{For five likely members with Gaia DR2 radial velocity data ($V_{rad}$)~km s$^{-1}$) (Col.~8), Gaia DR2 proper motion components $mas~yr^{-1}$ (Cols.~6 -- 7) and parallaxes (mas) (Col.~9). Their equatorial coordinates (J2000) (Cols.~2 -- 3) and $V$ and $(B-V)$ (Cols.~4 -- 5). Probability membership value (last column).}
		\setlength{\tabcolsep}{0.15cm}
		\begin{tabular}{cccccccccc}
			\hline
			STAR-ID & RA & DEC & $V$ & $(B-V)$ &$\mu_{\alpha}\pm\sigma_{\mu\alpha}$ & $\mu_{\delta}\pm\sigma_{\mu\delta}$ &$V_{rad}\pm \sigma_V$ & $\varpi\pm\sigma_\varpi$& P($\%$) \\
			\hline
			508 &30.3497 &75.5020 &12.932 &2.226 &2.716 $\pm$ 0.064 &0.397 $\pm$ 0.066 &-28.05 $\pm$ 3.20 &0.239 $\pm$ 0.040 & 99.77 \\
			862 &30.1545 &75.4700 &13.936 &2.165 &2.476 $\pm$ 0.048 &0.420 $\pm$ 0.051 &-26.40 $\pm$ 0.78 &0.257 $\pm$ 0.032 &99.66 \\
			737 &30.2315 &75.4944 &14.520 &1.918 &2.643 $\pm$ 0.040 &0.484 $\pm$ 0.039 &-28.55 $\pm$ 0.84 &0.302 $\pm$ 0.024 &99.73 \\
			299 &30.5162 &75.4812 &14.615 &1.956 &2.483 $\pm$ 0.041 &0.500 $\pm$ 0.041 &-30.88 $\pm$ 0.92 &0.273 $\pm$ 0.025 &99.67 \\
			674 &30.2739 &75.4492 &14.963 &1.707 &2.588 $\pm$ 0.036 &0.261 $\pm$ 0.038 &-31.57 $\pm$ 3.86 &0.235 $\pm$ 0.023 &99.67 \\
			\hline
		\end{tabular} 
	\end{center} 
}
\end{table*}

By utilising the "MWPotential2014" code in the galpy-code library \footnote[1]{http://github.com/jobovy/galpy [accessed 20 May 2019]} written by Bovy (2015) \cite{bov15}, peri- and apo-galactic distances $(R_\text{min},~R_\text{max})$ (kpc) and the maximum height distance (z$_{max}$) (kpc) have been obtained. The orbital eccentricity (ecc) is estimated via the relation $e = (R_\text{max}-R_\text{min})/(R_\text{max}+R_\text{min})$.  Mean orbital radius ($R_{m}$) (kpc) is given as the mean of $R_\text{min}$ and $R_\text{max}$ distances. Each member's orbit has been integrated for 2.8 Gyr (Table 4) within the Galactic potential. The galactic potential is a sum of the Galactic components, as given in the paper of Bovy (2015) \cite{bov15}.  

\renewcommand{\arraystretch}{1.4} 
\begin{table*}[!h]\label{Table-6}
	{\small
		\begin {center}
		\caption{For five probable members of Be~8, kinematics ($U$, $V$, $W$, $V_{\Phi}$) km s$^{-1}$ (Cols.~2--5) and orbital parameters (R$_{max}$, R$_{min}$, R$_{m}$, z$_{max}$ (kpc)) and ecc (Cols.~6--10). Their orbital angular momentum values ($J_{z}$ and $J_{\bot}$) (kpc km s$^{-1}$) (Cols.~11 -- 12).}
		\setlength{\tabcolsep}{0.28cm}
		\begin{tabular}{cccccccccccc}
		\hline
		STAR-ID & $U$ & $V$ &$W$ &$V_{\Phi}$& R$_{max}$ & R$_{min}$&R$_{m}$ & z$_{max}$& ecc  &$J_{z}$ & $J_{\bot}$ \\
		\hline
		508 &-19.77 &-55.89 &15.36 &-184.42 &12.44 &6.69 &9.56&1.27 &0.30&-2056.63 &259.33 \\
		862 &-14.94 &-49.42 &13.32 &-192.93 &12.12 &7.09 &9.60&1.15 &0.26&-2106.96 &242.69 \\
		737 &-10.64 &-48.43 &11.61 &-196.75 &11.32 &7.10 &9.21&0.97 &0.23&-2060.68 &218.84 \\
		299 & -9.91 &-51.27 &12.69 &-193.21 &11.66 &7.07 &9.37&1.07 &0.24&-2076.34 &235.59 \\
		674 &-17.44 &-57.54 &11.64 &-183.29 &12.40 &6.68 &9.54&1.21 &0.30&-2054.53 &233.05 \\
		\hline
		\end{tabular}  
		\end{center}
	}
\end{table*} 	

The orbital angular momentum components $J_{x}$, $J_{y}$, $J_{z}$ and J$_{\bot}$ (kpc km s$^{-1}$) for five members are calculated from the equations of Kepley \etal (2007) \cite{kep07}. These orbital and angular momentum parameters ($J_{z}$,~J$_{\bot}$) are given in Cols.~6--12 of Table 6.
The total angular momentum J$_{\bot}$ is defined as $J_{\bot}=(J_{x}^2+J_{y}^2)^{1/2}$.  For example,  the right handed J$_{z}$ value of a star near the Sun is $-$1960 kpc km s$^{-1}$ from the solar values of R$_{\odot}$ and $V_{LSR}$. 

The vertical heliocentric velocities of W $=$ [12, 15]  km s$^{-1}$ and azimuthal velocities,  $V_{\Phi}=[-183,- 197]$ km s$^{-1}$ of these five likely members indicate that they have typical Galactic disk velocities. They reside in outer Galactic disk, $R_{m} > 8.5$ kpc with the circular orbits, ecc $=$ [0.23,~0.30]. Their Galactic heights reach to  $z_{max}=1.27$~kpc . According to Figure 13 of Carney \etal (1996) \cite{car96}, thin and thick disk stars have the orbital eccentricities with $ecc < [0.25,~0.30]$ (circular orbits) and  $0.30 < ecc < 0.45$ (elliptical). In the sense their orbital parameters reflect the properties of the Galactic thin disc.  This is also consistent with what is expected of its metal content, [M/H] $=$ -0.27.

The Galactic orbits of five probable members of Be~8 are presented in Figure 11(a) and (b).  Five likely members with the circular orbits move from $\sim$12 kpc to $\sim$7 kpc on x -- y (kpc) (panel a). On z -- R (kpc) (panel b), they reach to z $\sim$1.3 kpc for the range of $6 < R < 12$ kpc. Here R means the Galactocentric distance. The angular momentums (J$_{\bot}$,~ J$_{z}$) of these members fall in the range of  $220 < J_{\bot} < 262$ kpc~km~s$^{-1}$ for $-2107 \le J_{z} \le -2054$ kpc~km~s$^{-1}$. Their kinematics, dynamical and angular momentum values imply the indicators of Galactic thin disk population, according to fig. 11 of Kepley \etal (2007) \cite{kep07}. 

\begin{figure}[!h]\label{Fig-11}
	\begin {center}
	\includegraphics[width=0.41\columnwidth]{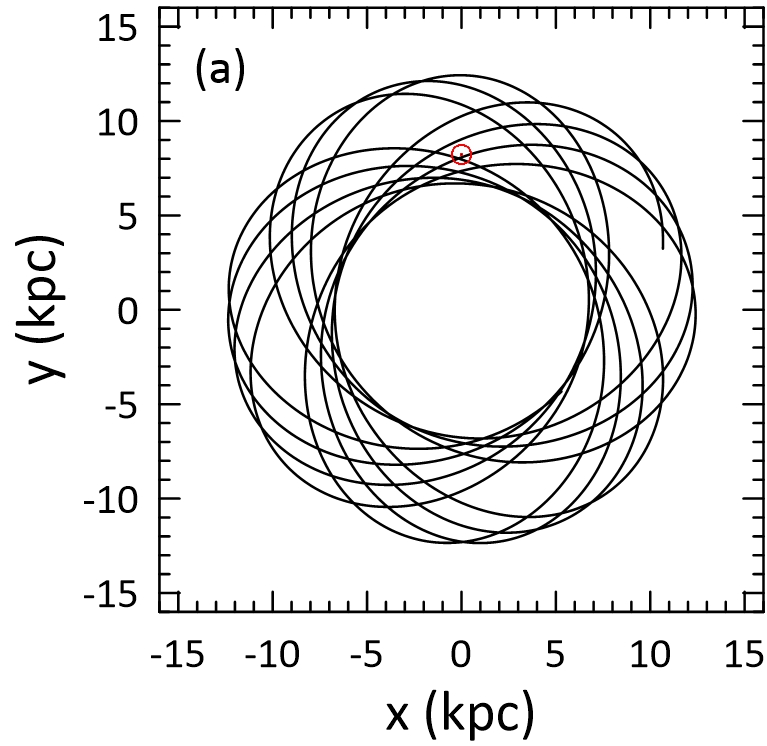}\hspace{1.2cm}
	\includegraphics[width=0.40\columnwidth]{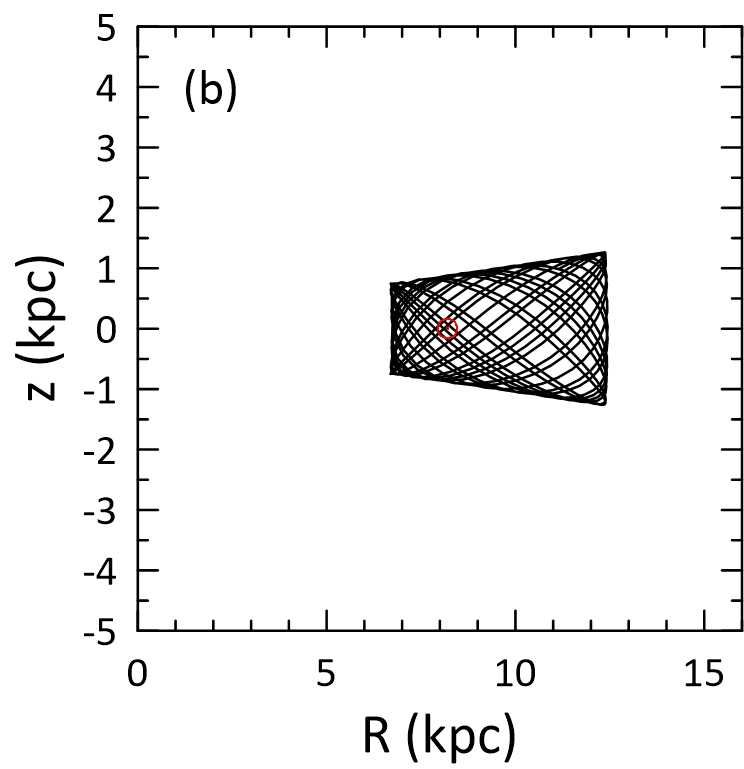}
	\caption{The orbits of five probable members of Be~8 on x -- y (kpc) (panel a)  and z -- R (kpc) (panel b). 
		Large circle denotes the position of the Sun, $(z_{\odot},~R_{\odot})=(0,~8.2~kpc)$.}
	\end {center}
\end{figure}

\section{Discussions and Conclusions}

For the cluster members of Be~8, the reddenings E($B-V$) = 0.69 $\pm$ 0.03 and E($V-I$) = 0.87 $\pm$ 0.10 have been obtained from $V$~-~$(B-V)$ and $V$~-~$(V-I)$ diagrams (Figure 8). These values are in good concordant with the ones of E($B-V$) $=$ 0.75 and E($V-I$) $=$ 0.88, given by Hasegawa \etal (2004) \cite{has04}. Our E$(B-V)$ value is larger than 0.41 value given by Bukowiecki \etal (2001) \cite{buk11} for 2MASS JHK$_{s}$ photometry.
	
Our distance modulus and distance from $V$~-~$(B-V)$ are ($V_{0}$~-~$M_{V}$,~d(pc)) = (12.66 $\pm$ 0.19, 3410~$\pm$~300 pc), which are in good agreement with the value (12.49, 3148~pc) of Hasegawa \etal (2004) \cite{has04}. Our value is somewhat smaller than the value (13.34,~ 3960~pc) of Bukowiecki \etal (2001) \cite{buk11}. Our distance modulus/distance (pc) and age (Gyr) from $G-(G_{BP}$~-~$G_{RP})$ are in quite compatible with the ones of the colour indices, $(B-V)$, $(V-I)$ and $(R-I)$ (Table 4). The median Gaia DR2 parallax from 43 likely members ($\sigma_{\varpi}/\varpi$ $<$ 0.20) gives us $\varpi$ = 0.272 $\pm$ 0.060 mas which corresponds to d $=$ 3676 $\pm$ 810 pc.  This distance is in good concordance  with the photometric distances (3410--3620 pc) within the uncertainties (Table 4). A global systematic offset of Gaia DR2 parallaxes is $\Delta\varpi$ $=$ $-$0.029 mas in terms of an inertial reference frame, derived by Lindegren \etal (2018) \cite{lin18}. Recent values for the zero point shift of parallax have been found as  $\Delta\varpi =-$0.045 $\pm$ 0.009 mas \cite{yal18}, $\Delta\varpi = -$0.053 $\pm$ 0.003 mas \cite{zin18}, and  $\Delta\varpi = -$0.046 $\pm$ 0.013 mas \cite{rie18}, respectively.A correction of 0.005 mas to the median value of our median parallax gives a closer distance with a difference of 66 pc for Be~8. 

The age of Be~8 is 2.80 $\pm$ 0.20 GYr and this is younger than 3.16 GYr of Hasegawa \etal (2004) \cite{has04} and Bukowiecki \etal (2001) \cite{buk11}. Its MAI age (A $=$ 2.94 Gyr) is in concordant with its isochrone age. Discrepancies of the distance moduli, distances and ages as compared to the literature stem from the usage of different heavy element abundances, isochrones,  reddenings and photometries such as 2MASS JHK$_{s}$, as mentioned by Moitinho (2010) \cite{moi10}.	

Two BS candidates reside in core radius of 1.8$^{\prime}$ of Be~8 (Figure 10). BS stars potentially locate in the inner regions of stellar clusters \cite{car01}. According to Ferraro (2016) \cite{fer16}, their formation ways are explained: mass transfer in binary systems \cite{mc64} due to the merging of the two stars and stellar collisions \cite{hil76}. 

It is surprising to find Be~8 with $[M/H]=-0.27$ (close to solar metallicity) at such large galactic radius (R $=$ 10.57 kpc). However, the orbits in Fig.11(a) and (b) show that the cluster passed a part of its time at galactocentric radius, $R=6-7$ kpc, and then possibly it was born at that radius, which would explain the metallicity. According to Figure 3 -- 4 ($[M/H]$ versus R~(kpc)) of Lepine \etal (2011) \cite{lep11}, Be 8 with $[M/H]=-0.27$ and R $=$ 10.57 kpc reside in a region of R $>$ 9 kpc (co-rotation gap at 9 kpc). In the sense it may have been originating from different galactic radius or different star formation region \cite{lep11}.

\section*{Acknowledgements}
We wish to thank the staff of Sierra Nevada Observatory, Granada-Spain. RM acknowledge the financial support from the UNAM under DGAPA grant PAPIIT IN100918. We thank J. Lepine for his valuable comments. This paper has made use of results from the European Space Agency (ESA) space mission Gaia, the data from which were processed by the Gaia Data Processing and Analysis Consortium (DPAC). Funding for the DPAC has been provided by national institutions, in particular the institutions participating in the Gaia Multilateral Agreement. The website of Gaia mission is http://www.cosmos.esa.int/gaia.

\end{document}